\documentclass[letterpaper,twocolumn,10pt]{article}
\pdfoutput=1
\usepackage[pdftex]{graphicx}
\usepackage[intlimits]{amsmath}
\usepackage{amsfonts,amssymb}
\DeclareSymbolFontAlphabet{\mathbb}{AMSb}
\usepackage{float}
\usepackage[bf]{caption}       
\setcaptionmargin{0.5in}
\usepackage{fancyhdr}
\usepackage{fancyheadings}
\usepackage{fancybox}
\usepackage{ifthen}
\usepackage{url}
\usepackage{lscape,afterpage}
\usepackage{xspace}
\usepackage[FIGTOPCAP]{subfigure}
\usepackage[T1]{fontenc}
\usepackage[inline]{enumitem}
\usepackage[nocompress]{cite}
\usepackage[normalem]{ulem}
\usepackage{array}
\usepackage{booktabs}
\usepackage{color}
\usepackage{datetime}
\usepackage{epsfig}
\usepackage{inconsolata}
\usepackage{listings}
\usepackage{lscape,afterpage}
\usepackage{pifont}
\usepackage{times}
\usepackage{xcolor,colortbl}
\usepackage[binary-units=true]{siunitx}
\usepackage[utf8]{inputenc}
\usepackage[autostyle]{csquotes}
\usepackage{wrapfig}
\usepackage{nameref}
\usepackage{varioref}
\usepackage{hyperref}
\usepackage{cleveref}
\usepackage{epigraph}
\usepackage{tcolorbox}

\tcbset{colframe=black!75!black,colback=white,title=Observations on Elasticity, enlarge top by=1cm,enlarge bottom by=1cm}
\DeclareSIUnit\bps{bps}

\definecolor{mygreen}{rgb}{0,0.4,0}
\definecolor{mygray}{rgb}{0.5,0.5,0.5}
\definecolor{mymauve}{rgb}{0.58,0,0.82}
\newcommand{\ec}{UC}
\newcommand{\ecs}{UCs}

\newcommand{\stk}{Snapshot Stack}
\newcommand{\stks}{Snapshot Stacks}

\newcommand{\snap}{Snapshot}
\newcommand{\snaps}{Snapshots}

\newcommand{\spawning}{deploying}

\newcommand{\Spawning}{Deploying}

\begin{document}

\title{SEUSS: Rapid serverless deployment using environment snapshots}
\author{
  {\rm James Cadden, Thomas Unger, Yara Awad, Han Dong, Orran Krieger, Jonathan Appavoo}\\
  Department of Computer Science, Boston University \\
  {\textit{\{jmcadden,tommyu,awadyn,handong,okrieg,jappavoo\}@bu.edu}}
}
\date{} 
\maketitle

\thispagestyle{empty}

\label{ch:seuss}

\section{Introduction}
Serverless computing has been heralded as a \textit{general-purpose} programming model that enables easy access to on-demand cloud computation~\cite{berkleyview}. This is best illustrated by the FaaS model~\footnote{Function-as-a-Service}, wherein application logic is composed of short, high-level scripts (aka. \textit{serverless functions}) that are deployed automatically by a remote FaaS platform in response to demand. In serverless functions, application developers find a fine-grain programming primitive that is easy-to-use, naturally parallelizable, and can grow and shrink across arbitrary scales.

Work remains to bring serverless computing out of its current computational niche and toward a larger superset of \textit{general computing}. We imagine a FaaS platform wherein functions start as quickly and pack as densely as processes. To achieve this, we assert that FaaS platform must be able to support two broad categories of computation. First, the common case of repeatable, loop driven sequences and fixed degree parallelism. Second, \textit{elastic} computation characterized by request bursts, large-scale parallel fanouts, invocation diversity, and singleton execution.

Modern FaaS systems perform well in the case of repeat executions when
function working sets stay small. However, these platforms are less effective when applied to more complex, large-scale and dynamic workloads~\cite{Malawski2016, Kablan2017}. At the root of the problem is a \textit{lack of elasticity} in the underlying operating system mechanism used to deploy serverless function executions~\cite{KollerDan}. While great for maintaining individualized execution state for rapid re-invocation of functions, these coarse-grain isolation mechanisms are slow to construct, have large memory footprints, and are burdened by legacy software bottlenecks.

In this paper, we introduce a new system-level approach for rapidly deploying serverless functions. Through our approach, we demonstrate orders of magnitude improvements in both function start times, as well as in function cacheability, which improves common case re-execution paths while also unlocking previously-unsupported elastic FaaS workloads. 

Our model, SEUSS, or \textit{serverless execution via unikernel snapshot stacks}, combines  the following three OS techniques to create a novel approach  for deploying isolated function executions: 

\begin{enumerate} 
\item Functions are deployed within \textit{unikernels} equipped with a
  high-level language interpreter and a POSIX-like environment with common
    features like networking and a filesystem. Unikernels provide us with
    strong isolation of untrusted guest execution~\cite{Williams:2018}.
    Furthermore, the unikernel serializes the execution environment into a
    single flat address space, enabling our use of \textit{unikernel snapshots}.  
\item Initialized unikernel environments are captured in-memory as \textit{snapshot} images, which are then used as templates to construct additional environments for new function executions. We use pre-initialized environments to dramatically reduce the start time for deploying functions.
\item Unikernel environments undergo aggressive sharing of system state which drastically reduces the memory footprint of snapshot images and unikernel instances, enabling a high-density caching strategy for fast re-deployment of function executions from pre-initialized state.
\end{enumerate}

Unikernels have been previously explored as a potential solution for rapid serverles deployments~\cite{myvm,KollerDan}. In these cases, the unikernels had tiny memory footprints and boot in milliseconds, but only supported a highly-restricted execution model. In contrast, a unikernel that supports a POSIX-like environment with a full-featured language interpreter is both larger (100+ Mb) and slower to boot (~300 Ms). Thus, it is through \textit{unikernel snapshots} that SEUSS achieves its advantage. Furthermore, our design reveals a surprising contribution: it is possible to enable a high-performance FaaS backend from general-purpose software combined with a simple set of OS techniques.

To demonstrate the advantages of our approach, we have implemented SEUSS within a prototype kernel for virtualized x86 environments. When compared with the standard approach of deploying functions inside of Linux containers, SEUSS demonstrates: 
\begin{enumerate*} 
\item 50x reduction in the minimal deployment time for an uncached function
\item 17x more concurrent function environments active on a node
\end{enumerate*} 
When our SEUSS prototype is used as a drop-in replacement for Linux in a
multi-node Apache OpenWhisk deployment, we demonstrate:
\begin{enumerate*} 
  \item $51\times$ better platform throughput when running uncached workloads 
  \item Support for large-scale invocation \textit{bursts} that overwhelm the unmodified platform 
\end{enumerate*}

\begin{figure}[h]
\centering
\includegraphics[width=\columnwidth]{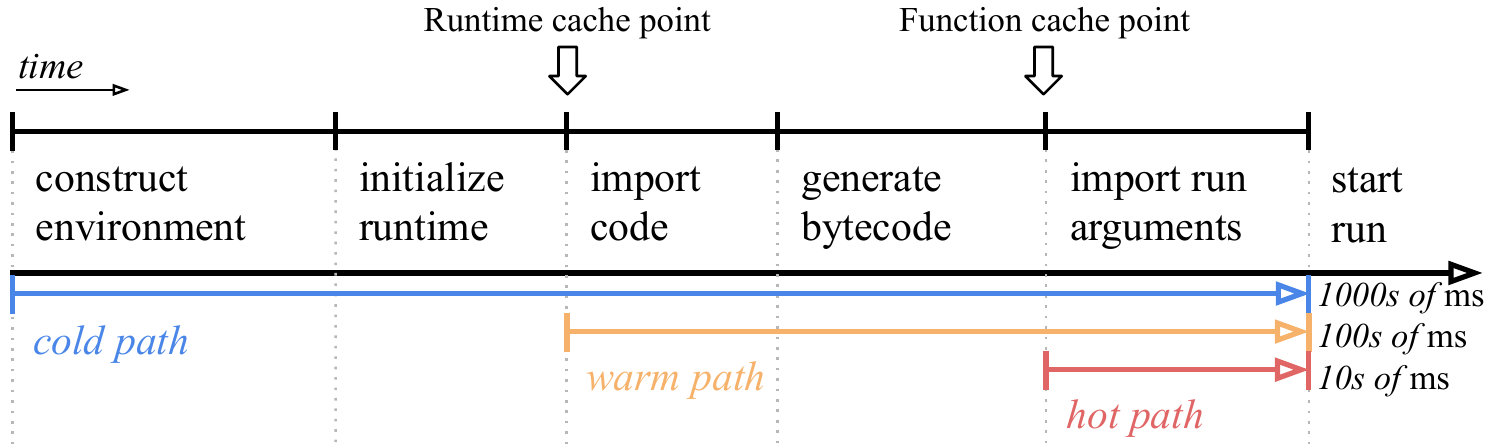}
\caption{Stages of a function invocation} 
\label{fig:invocation}
\end{figure}

\section{Background}
\label{ch:background}

In the FaaS model, clients upload the source code of their functions to the FaaS platform, and, in return, the client is given a handle used to signal a run of the function on a given set of inputs.  Once signaled, the FaaS platform is responsible for quickly scheduling and deploying the corresponding function code, a process we call a \textit{function invocation} (Figure~\ref{fig:invocation}).  To ensure safe execution on a multi-tenant FaaS platform, function executions take place inside of isolated containers or light-weight virtual machines instances which are typically dedicated to the function or client application~\cite{berkleyview}.

A general technique taken by FaaS platforms to improve function start times is to maintain \textit{caches} of initialized runtime environments (i.e. pools of idle VMs or containers) that sit ready for use in an upcoming function deployment. From this perspective, an invocation \textit{cold start} is effectively a cache miss, as it occurs when no environment exists ready for use in the invocation. In a cold start, the complete set of initialization steps are required before the function can start running. This adds multiple seconds to the deployment time as a new container or VM instance must be created. 

Comparably, an invocation \textit{warm start} is a cache hit on an isolated runtime environment ready for the function sources to be imported. In the case of interpreted runtimes like Node.js and Python, an intermediate compilation step is required to process the imported source code into executable bytecode prior to it being run (adding further overheads to the invocation paths). The overheads of importing and compiling the source code are specific to the function being invoked. However, even a simple “hello world” function is likely to add a few 100s of milliseconds in overheads (and possibly much higher for functions which import complex libraries~\cite{SAND}). 

Thus, the fastest and most optimal \textit{hot start} invocation is enabled using an environment that has been fully-initialized with the runtime and interpreted bytecode that is specific to the function being invoked. This can be accomplished by caching idle environments after they’ve been used for a function execution, such that they may be reused across future invocations of that given function. In this case, a new set of input arguments can be imported into the environment, and the function can start running in a few milliseconds (or less).

\section{Motivation}
\label{ch:motivation}

To better understand the opportunities that exist for specialization at the system level, we next highlight three characteristics of the serverless/FaaS model that simplify the role of the OS and unlock new opportunities for performance optimizations in function start times and high-density caching of function state. 

First, the operational shift from the application developer on to the remote FaaS platform enables the platform to adopt new and highly-specialized approaches for how serverless functions are deployed and isolated. For example, serverless deployment strategies have been proposed that use light-weight virtualization~\cite{myvm,gvisor}, light-weight containers~\cite{SOCK}, application-level isolation~\cite{SAND}, and language-level isolation~cite{cloudfair}.

Second, in the serverless model, function executions are defined to be independent, transient, and stateless. For example, there is no implicit sharing across parallel executions, and local modifications are assumed valid for only as long as the particular execution is alive. This simplified execution model avoids many of the features (and much of the complexity) provided by standard operating system mechanisms. Instead, it introduces the opportunity for designing simplified OS mechanisms that focus entirely on the rapid deployment of independent, short-lived executions.

Third, the FaaS platform defines a small set of language runtimes that function code must be written to. This can be seen by modern FaaS platforms that maintain “warm pools” of idle VMs that have been pre-initialized with a particular runtime environment~\cite{peeking}. A limited number of supported runtimes means that all function execution is derived from a small set of base environments and configurations. The implication of this is that a larger amount of state is likely to be identical across parallel executions of functions. Thus, an opportunity exists for fine-grain sharing of kernel state, file systems state (e.g. runtime binaries and shared libraries), and all the way down to the pages of the runtime processes themselves. 

In SEUSS, we exploit these three opportunities in a new deployment strategy designed to dramatically shorten function start times: \textit{serverless execution via unikernel snapshots}. The use of unikernels provides us with a strong isolation technique together with a simple self-contained abstraction, thereby enabling new performant techniques for the capture and deployment of unikernel snapshots. Furthermore, the single-address space representation of unikernels allows us to make extensive use of sharing between similar environments, thus decreasing the memory footprints of cached functions state.

\section{Related Work}
\label{ch:related}

SEUSS is not alone in arguing for system-level solutions towards improved FaaS performance.  In particular, SOCK~\cite{SOCK} and SAND~\cite{SAND} similarly strive for faster start times and improved cacheability through the use of light-weight isolation strategies and copy-on-write sharing across parallel executions.  What sets SEUSS apart from these similarly-acronym'd systems is our attempt at a general solution that is readily applicable across language runtimes and function invocation types. For example, deploying from snapshots can be applied to any environment run within a unikernel. In addition, sharing across unikernel snapshots provides an approach for page saving that is more comprehensive than that of forking processes.   

The underlying techniques of SEUSS have similarities (and applicability) to
use cases beyond serverless computing.  For example, edge computing benefits
from the safe, low-latency, event-triggered deployment of
unikernels~\cite{jitsu}. Execution snapshotting has been previously explored
in the context of process migration~\cite{criu}, fast
reinitialization~\cite{Wang:1995:CA:874064.875647}, and replay
debugging~\cite{LMC}. The isolation properties of a libraryOS deployed behind
a narrow domain interface is a technique shown to be effective against various
threat models~\cite{denali,drawbridge,haven,KylinX,gvisor,embassies}. 

The system-level approach of SEUSS has been motivated by previous cloud research systems that tailor their design to the characteristics of an application, with the goal of significantly increasing platform utilization and throughput~\cite{potemkin,snowflock,kaleidoscope,denali}. In particular, our use of page-level sharing via snapshot stacks is most similar to page-level sharing between VMs in Potemkin~\cite{potemkin}.  Much akin to the
distributed paging techniques employed by Snowflock~\cite{snowflock} and
Kaleidoscope~\cite{kaleidoscope}, we view the natural evolution of our cache model
is to expand across a cluster of compute nodes, thus introducing
\textit{distribution and replication} to SEUSS\footnote{At which point we will
be obliged to rename it to \textit{DR-SEUSS}}.

\section{SEUSS Execution Model}
\label{ch:seussexec}

Three techniques underpin the SEUSS execution model. \textit{\ecs} are used to circumscribe all necessary function state into a flat address space. \textit{\snaps} are used to capture and deploy functions at arbitrary points during their execution. \snaps\ enable fast function deployment and amortize deterministic startup. The resulting fast function start times introduces elasticity into the FaaS platform, especially for the execution of uncached functions. Furthermore, \stks\ enable fine-grain memory sharing between \snaps\ with shared lineages. This dramatically increases the FaaS platform’s ability to cache function-specific environments, ready for immediate reuse.

\subsection{Unikernel Contexts (\ecs)}
\label{ch:seussexec:execcontext}
Most every operating system has a representation for the state of a running program; \ecs\ comprise our system abstraction for isolating running functions as manageable system objects. \ecs\ are wrappers for: 1) the unikernels that hold function state as well as 2) a small set of system \textit{meta-state} for handling them.

Unikernels are natively executable binaries that contain an application, its execution dependencies, and a kernel-level support library (also called a \textit{LibraryOS}), all linked together into a single address space. For use in for FaaS executions, every \ec\ instance contains a unikernel equipped with full-feature language runtime environment. As opposed to minimal-footprint or domain-specific unikernel, we chose to adopt a feature-rich unikernel that provides a POSIX-like environment and a common set of features like full network stack, shared libraries, and a root filesystem. This software stack is not optimized for speed or size---in fact, it takes hundreds of milliseconds to boot a Node.js unikernel ready to receive input over the network---but performance is won back through our use of \textit{unikernel snapshots}.

A critical advantage of a unikernel approach is that its address space contains the vast majority of the mechanisms required for executing arbitrary function code. Thus, the amount of execution-specific state that must be managed \textit{outside} the unikernel is minimal. The implication of this is a major simplification for snapshotting: the function state \textit{is} the address space of the unikernel. There is no marshalling, quiescing, or synchronizing around in-kernel data structures, enabling snapshotting and deployments to be sub-millisecond operations (\S\ref{ch:evaluation:microbenchmarks:seuss}).

\subsection{Snapshots}
\label{ch:seussexec:snapshots}

A \snap\ is an immutable data object which captures a time point in the lifetime of an \ec.  From a traditional OS perspective, snapshots are read-only memory images packaged together with the register state and the system-level meta-state for an executing process.  In SEUSS, \snaps\ act as templates from with many addition \ecs\ can be deployed. By capturing snapshots at strategic points along commonly used code path, we significantly shorten the initialization steps required to deploy new execution. 

Furthermore, \snaps\ enable powerful \textit{anticipatory optimization}. This is done by snapshotting after speculatively executing code paths likely to be used during \ec\ execution. The goal here is to accumulate a significant amount of allocated state into the “base” snapshot, reducing the memory footprint of the \ecs\ branched from it. In our system, we use these techniques to exercise the network stack, the interpreter compiler, and the interpreter execution path.  By doing initialization and "warm-up" work prior to capturing the base \snap\, SEUSS significantly reduces the latency and memory required to store a \snap\ and to run an \ec\, while maintaining identical external effects.

\subsection{Snapshot Stacks}
\label{ch:seussexec:snapshotstacks}

In SEUSS, we optimize \snaps\ by \textit{factoring out} the common execution state into \stks\, enabling multiple \snaps\ and \ecs\ to be backed by the same physical memory pages. This dramatically decreases the memory footprint of \snaps\ and \ecs\ and increases the amount that can be stored on a single machine. \stks\ are enabled by capturing into a \snap\ only the pages modified since the \ec\ was created.  For this, we use traditional copy-on-write semantics enabled by hardware to track recently changed memory state. In this way, a \stks\ is an ordered sequence of \snaps\ with each entry acting as a page-level \textit{diff} of the previous. For example, a single heavyweight \snap\ of the JavaScript interpreter can be used as a base \snap\ with further \snaps\ containing only the function-specific memory state. 

To better understand the advantage of \stks, consider the following example: a FaaS platform needs snapshots to capture the fully-initialized state of JavaScript functions \texttt{Foo()} and \texttt{Bar()}. Armed with only a snapshot mechanism the platform requires two \ec\ snapshots, one for each function.  With \stk, three snapshots are required, one for the initialized JavaScript runtime unikernel, and one for each specific functions. The JavaScript snapshot, which is created ahead of time, is used to deploy \ecs\ for each of the two functions at the time of invocation. As these \ec\ move forward, it writes to memory accumulate a delta relative to the read-only state of base snapshot it was deployed from.  Therefore, when the time comes to capture the two function-specific snapshots, only \textit{ difference} between the base snapshot and the memory and the targeted \ec\ will be copied into the new \snap\ image. This \snap\ now constitutes a \stk, representing both the base \snap\ and function \snap.  

In our SUESS OS prototype, \stks\ account for a savings of over 100 Mb \textit{per snapshot} for functions deployed using a Node.js unikernel(\S~\ref{ch:evaluation:microbenchmarks}, table~\ref{table:unikernel_size}).

\subsection{SEUSS \textit{in action!}}
\label{ch:seussexec:action}
\begin{figure}
\centering
\includegraphics[width=0.92\linewidth]{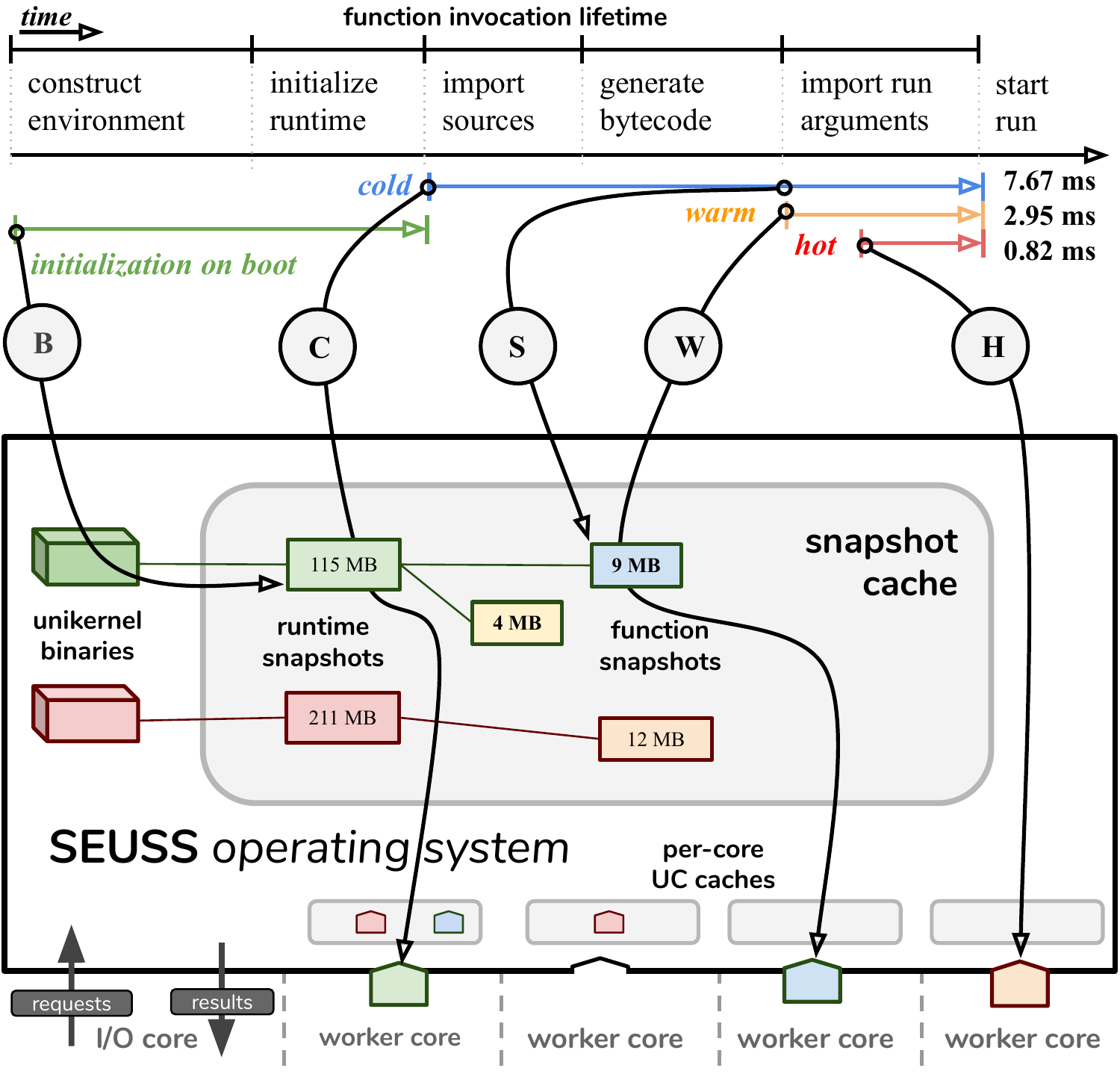}
\caption{
  The high-level procedures of SEUSS on a dedicated compute node.  }
\label{fig:seuss}
\end{figure}

We next describe the invocation procedures (hot, warm, and cold) for starting a serverless function using SEUSS. Figure~\ref{fig:seuss} depicts the high-level procedure of SEUSS on a dedicated compute node. Once received, the local system has a choice of three pathways to process the invocation, the choice of which is determined by the presence of cached state specific to the function being invoked. Cached state may be found in either the \textit{warm pool} or the \ec\ \textit{hot tub}.

\begin{description}
  \item[On SEUSS boot (“B” in Figure~\ref{fig:seuss})]
Runtime \snaps\ are taken for each interpreter during the system boot.  Each runtime \snap\ is of a runtime waiting to user source to be imported so that it can be compiled and run it on demand. 

  \item[Cold path (label “C” in Figure~\ref{fig:seuss})]
If no other cached function state exists, a runtime \snap\ is used to deploy an \ec. The function source is imported and compiled by the runtime. Once the compilation is finished, a function-specific \snap\ is created (“S” in the figure) and placed into the warm pool. Finally, the function run arguments are imported and the execution of the function begins. Once finished, this \ec\ can be destroyed, or optionally cached in the \ec\ hot tub for reuse across further invocations of that function. In our prototype, a cold start of a NOP JavaScript function finishes in 8ms.

  \item[Warm path (label “W” in Figure~\ref{fig:seuss})]
When function request hits in the warm pool cache, a new \ec\ is created from a function-specific \snap. In this case, compilation is skipped, run arguments are accepted, and execution begins. As in the cold path, the used \ec\ can then be deleted or optionally enter into the hot tub. The warm pool is comprised of immutable memory images (at most one per user function), from which any number of \ecs\ may be launched (at any degree of parallelism). Nothing is “consumed” when an \ec\ is created from a warm pool \snap. In our prototype, a warm start of a NOP JavaScript function finishes in 3ms.

  \item[Hot path (label “H” in Figure~\ref{fig:seuss})]
When a function request hits in the hot tub, an existing \ec\ is taken out of
    the hot tub, the run arguments are imported into the \ec\ and execution
    begins. The \ec\ may be re-entered into the hot tub upon completion. The
    hot tub works differently from the warm pool as the entries are “live”
    \ecs\, not immutable \snaps\. Thus, a cached \ec\ is consumed for the
    duration of a single execution. Further, \ecs\ may accumulate significant
    memory state after multiples usages, so naturally, the hot tub has shorter
    recommended occupancy than the warm pool. In our prototype, a hot start of a NOP JavaScript function finishes in less than 1ms.

\end{description}

\section{Implementation}
\label{ch:implementation}   

In this section, we describe our implementation of the SEUSS model within a prototype kernel designed to run on the compute nodes within a large-scale FaaS platform architecture. In addition to the implementation details of \ecs, \snaps, and \stks, we also describe our prototype solution for scalable \ec\ networking and how we integrate a SEUSS node into an Apache OpenWhisk cluster (for our evaluations in \S\ref{ch:evaluation_openwhisk}).

\subsubsection*{Overview}
\begin{enumerate}
\item The SEUSS OS is a light-weight multicore kernel that implements SEUSS in a native x86 virtualized environment.
\item To deploy new execution, SEUSS OS maintains a cache of \snaps\ as well as per-core caches of idle \ecs
\item The software stack of an \ec\ is implemented using the Rumprun unikernel linked together with an interpretive language runtime (Python or JavaScript) and invocation driver.
\item \ecs\ execute entirely in user mode (ring 3) with page-table based hardware protection on top of minimal domain interface.
\item A network layer masquerades traffic going in and out of \ecs\ allowing for outgoing network connections initialized from within the guest functions.
\end{enumerate}

\subsection{Operating System Design}
\label{ch:implementation:design}
Figure~\ref{fig:seuss} depicts the high-level operations of SEUSS within our purpose-build operating system, \textit{SEUSS OS}.  The low-level kernel manages the local resources of the node, which are partitioned across worker cores dedicated to running \ecs\ and IO cores are dedicated to external networking.
The low-level approach of the SEUSS OS demonstrates the ability to selectively apply OS-level specialization to an existing FaaS platform by targeting only the nodes where function execution takes place (and leaving everything else unchanged). In this way, with SEUSS OS it is possible to construct a high-density execution plane out of a standard set of virtual machines.

\subsubsection{System Boot}
As part of the SEUSS OS startup procedure, the base unikernels are booted from their ELF binary images and initialized. Next, a runtime snapshot is captured for each unikernel. At this point, the kernel registers itself with the FaaS platform and awaits for invocation requests to arrive. 

\subsubsection{Invocation Procedure}
Invocation requests are received from a remote FaaS controller through an IO core. Requests are then places onto a shared work queue that idle worker cores can pull from. Once a request is pulled by a core, it will check for cached state to accelerate the invocation, following the procedure outlined in \S\~ref{ch:seussexec:action}. 

Once an idle \ec\ is selected or a new \ec\ is constructed from a \snap, the worker then connects to the running \ec\ and issues commands. Initialized as part of the runtime bring up, the invocation driver deploys an HTTP/REST endpoint within the \ec\ that the worker connects to. This connection is used by the worker to import function code and arguments and to start the execution of the function.  Once the execution is finished, the results are passed to the worker and the connection is closed. The \ec\, which now contains the interpreted bytecode and imports generated during execution, is returned to a blocked listening state, allowing for the work to quickly re-execute the function with a new set of arguments (i.e. a hot start) 

Figure~\ref{fig:seuss} shows the sizes (in MB) of different snapshots images and
the relative deployment overheads of the invocation paths measured from within our prototype (full analysis is provided in \S\ref{ch:evaluation:microbenchmarks:snapshots}).

\subsubsection{Isolation \& Security}
Unikernels execute entirely in user mode (ring 3) with page-table based hardware protection. A highly-restricted domain interface sits between the \ec\ and the OS, which act as a 'narrow' attack surface that the untrusted \ec\ has access too. The isolation properties of a libraryOS deployed behind a narrow domain interface is a technique shown to be effective technique against various threat models~\cite{denali,drawbridge,haven,KylinX,gvisor,embassies}. 

\subsection{Software Stack}
A SEUSS x86\_64 prototype is designed to run natively inside of a \texttt{kvm-qemu} virtual machine.  The bottom-most layer the prototype uses the EbbRT LibraryOS framework\cite{ebbrt}, which provides a multicore event-driven runtime, bootstrapping logic, a \texttt{virtio} paravirtualized NIC, and a native TCP/IP network stack. On top of this foundation, we have built the components responsible for implementing the SEUSS execution model and enabling our use of cached \snaps\ and the safe deployment of functions in \ecs. The EbbRT framework has been extended to support for multiple protection modes via the \texttt{x86\_64} \texttt{sysret} and \texttt{syscall} instructions.

 \begin{figure}[th]
    \centering
 \includegraphics[width=0.9\linewidth]{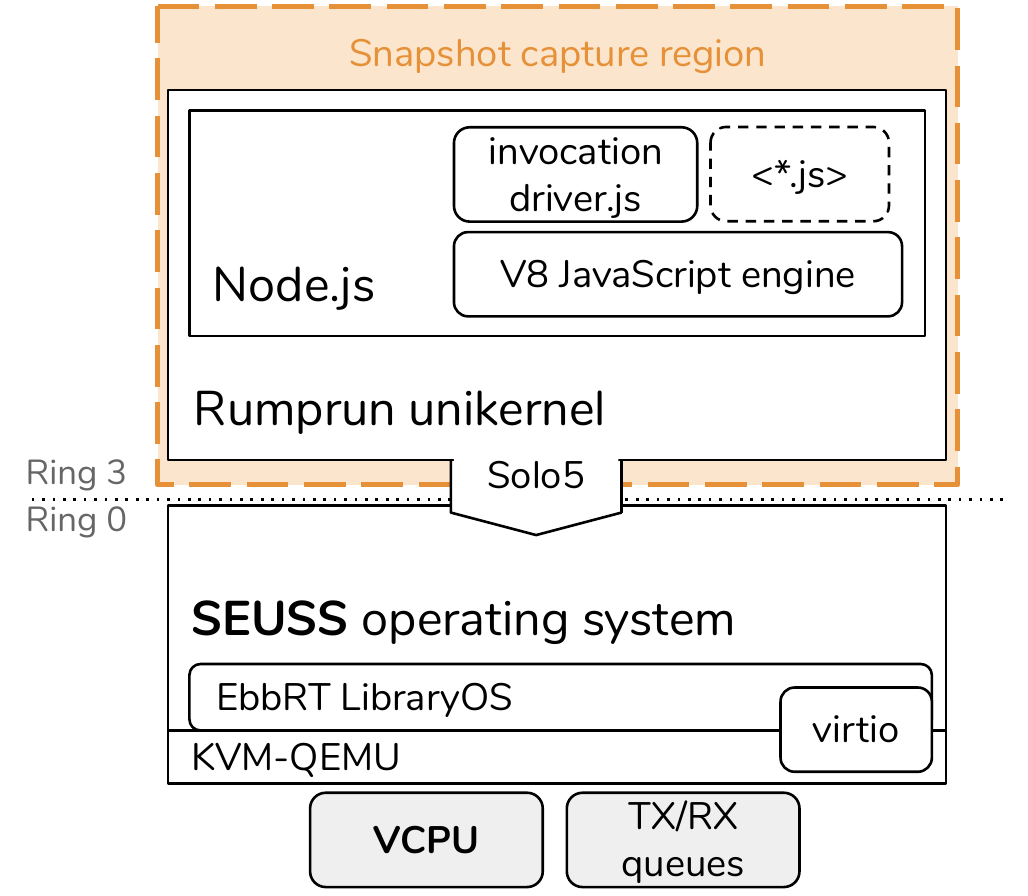}
  \caption{Vertical slice of the single-core system stack}
 \label{fig:stack}
  \vspace{-0.15in}
  \end{figure}

\label{ch:implementation:unikernel}

The SEUSS prototype uses the Rumprun unikernel linked with a port of Node.js or Python. The Rumprun unikernel provides a general-purpose POSIX-like execution environment based on the NetBSD kernel, which provides a common set of shared system libraries and a ramdisk filesystems~\cite{rumprun-git}. Rumprun is deployed on top of the \texttt{Solo5} unikernel monitor, which distills the low-level machine interface of the unikernel down to a minimal set of paravirtualized hypercalls used by a unikernel to access the outside world.  The minimal interface of \texttt{Solo5} was instrumental in enabling the simple and concise address space abstraction that acts as the foundation for \ecs\ and \snaps. 



\subsection{\snaps \& \stks}
\label{ch:system:snapshots}
Next, we describe our technique for implementing \snaps\ and \stks\. Our prototype uses direct access to hardware page tables to capture \snaps, deploy \ecs, and to enable fine-grain page-level sharing across “stacks” of \snaps\ and \ecs.

\subsubsection{Capturing \snaps}
\label{ch:implementation:snap:triggering} 
Crucial to our design was a mechanism to allow triggering the creation of a \snap\ from \textit{within the unikernel itself}. Doing so allows us to capture a \snap\ at precise moments specified within the high-level invocation driver source code. Language-level snapshot creation also opens the door open to future research into higher-level tooling that enables developers to directly instrument \snap\ points within their code, perhaps amortizing expensive data structure initialization.  

We make use of language-level snapshotting by capturing new snapshots immediately prior to when the invocation driver blocks on the listening socket. When we deploy new \ec\ from a snapshot, the invocation driver begins ready to accept a connection over a local network port.  

In our SEUSS prototype, \snap\ trigger is an expedient hack involving an \texttt{x86} debug register, which enables us to trigger a hardware exception prior to the execution of an instruction on a preconfigured linear address (which corresponds to a designated symbol within the language runtime included in the unikernel).  When the exception occurs, execution switches into kernel mode and calls into a kernel handler that records the state of the \ec\ into a new \snap. When finished, execution transitions back into user mode and continues from the instruction where the exception was triggered, entirely transparent to the running \ec.

\subsubsection{\Spawning\ \ecs}
\label{ch:implementation:snap:deploy}
The procedure of \spawning\ function execution from a \snap\ starts with creating a new \ec.  The TLB is flushed and the root of the \ec\ page table structure is mapped to the core.  Once this is complete, execution switches into the new \ec\ by triggering a breakpoint exception and overwriting the exception frame with the registers corresponding register values from the source snapshot. The interrupt service routine does the work of popping the registers back onto the core and execution transfers to the exact instruction where the snapshot was captured. 

\begin{figure}
\centering
\includegraphics[width=\linewidth]{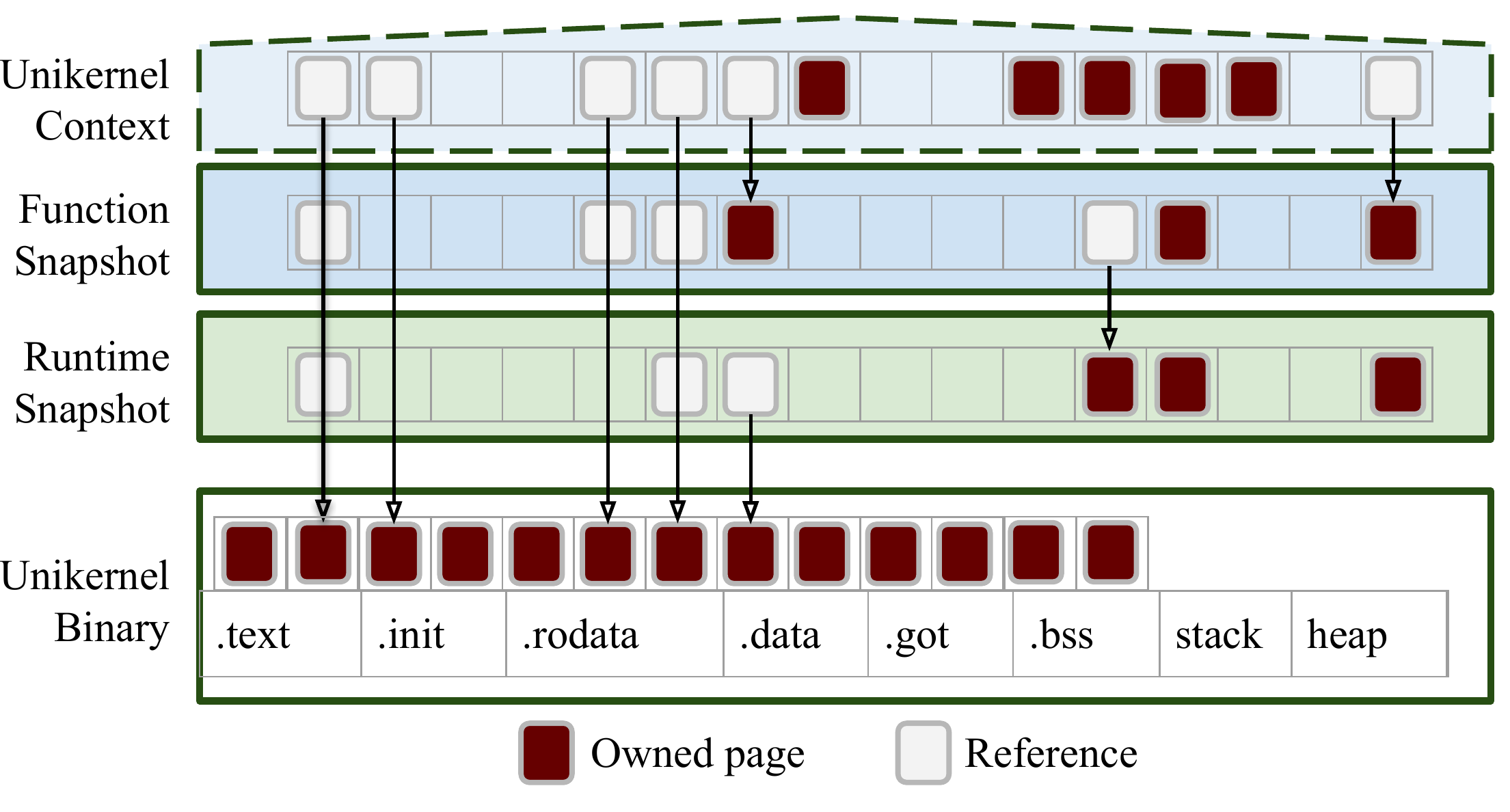}
\caption{
  Page sharing across \snaps\ and \ecs.
}
\label{fig:snapshot_stacks}
\end{figure}

\subsubsection{\stks}
\label{ch:implementation:snap:stacked} 

\snaps\ are kept light-weight by only capturing the pages modified since the \ec\ was created. In our model,a \stks\ is an ordered sequence of \snaps\ with each entry acting as a page-level diff of the previous (Figure~\ref{fig:snapshot_stacks})..    To achieve this, we use traditional copy-on-write semantics enabled by hardware to track recently changed memory state. This is accomplished by capturing the complete page table structure in each snapshot, but cloning only the pages that have been recently \textit{written} (designated on x86 with the \texttt{Dirty} bit). In addition, snapshots record the location of the top-level \snap\ or unikernel binary which the captured \ec\ was spawned from, which is then used to resolve fault during execution of an \ec.

When a new \ec\ is created, the procedure requires cloning the page table structure from within the snapshot. All the \texttt{Dirty} bit set for all of table entries, and the \texttt{Read}/\texttt{Write} bits are unset, enabling for copy-on-write faults.  Page fault that occurs within a \ec\ are processed by a kernel fault handler that is aware of the \stk\ backing the \ec\.
Depending on the semantics of the fault, the kernel handler may allocate a new page, clone a page from within the backing \stk, or resolve the fault with a read-only mapping to a page within the source \stk.  \footnote{Given these semantics, a \snap\ can only be deleted when it is known that no other \snaps\ or \ecs\ depend on it (e.g., a reference count). We avoid this concern in our prototype by only deleting the \textit{function-specific} snapshots that have no corresponding \ecs.}
o
\subsubsection{Memory Management}
\label{ch:implementation:memory} 
One issue with extensive use of copy-on-write is that memory becomes highly overcommitted and application demand may cause the system to run out of physical memory.  This is a problem, for example, with the heavy use of fork() in Linux, whereby an OOM daemon may be triggered and may kill system critical processes.  This is not a problem in our design, because \ecs\ for function invocations are transient and can always be killed by the system without impacting the system’s ability to make forward progress.  While more complexity may be necessary in the future, our OOM daemon for SEUSS is trivial: it reclaims idle \ecs\ that do not currently host a live invocation as soon as the available physical memory drops a pre-defined some threshold.

\subsection{\ec\ Networking}
\label{ch:implementation:networking}
Each \ec\ is configured with an identical IP and MAC address, thus enabling \snaps\ to be trivial deployed across time, in parallel across cores, and enables the potential for us to migrate \snaps\ across machines.  A network layer monitors traffic going in and out of the \ec\ and enables internal and external communication by forwarding and masquerading traffic down through the SEUSS network stack.  The internal network allows the SEUSSinvocation procedure to communicate with the running unikernel, for example, to send the input arguments to the invocation driver. The external network proxy monitors the unikernels access to the outside world.  

A per-core \textit{network proxy} maintains mappings for both the internal and external networks, for each unikernel instance active on that core and routes.   Incoming traffic is screened, and the traffic destined for unikernels sent through an additional translation process to determine the worker core where the \ec\ is resident.  TCP destination ports act as the unique key for mapping packets to an active \ec.  We currently do not support port mapping of UDP or IPv6 packets, but the approach would be similar.  This design only supports outgoing TCP connections initiated from within the unikernel. To support listening sockets within the unikernel would require some out-of-band communication between the unikernel and the network proxy to signal intent, but due to the transient nature of serverless functions we have yet to give this much thought. 

\subsection{OpenWhisk Integration}
\label{ch:seuss:implementation:openwhisk} 
OpenWhisk provides us with a modern, full-feature FaaS computing platform, which provides many functions beyond the invocation procedures target by SEUSS. For example, OpenWhisk includes management and authentication of users, uploading and editing of functions, the enforcement of usage quotas, whitelist \& blacklists, long term storage, etc. To preserve this important platform functionality, we have designed SEUSS prototype kernel to act as a protocol-compliant "drop in" replacement for the OpenWhisk compute nodes, while all other platform functionality remains unaffected. 

To accomplish this, we have built an intermediate shim process, written in C++ and run on Linux, responsible for reading request off the OpenWhisk message bus (Kafka), translating each request to an internal message which is sent from the shim process to the SEUSS VM deployed on the same machine.  The advantage of this approach is that it avoids the need to support a  platform-specific protocol within our lightweight OS runtime. Instead, the shim processes can be written in any language to take advantage of the wealth of existing client libraries for the services that make up the FaaS platform (e.g., Kafka, CouchDB, ZooKeeper). In our prototype, our shim process registers itself as an OpenWhisk invoker and received action requests through the Kafka message bus. Function code is read out from CouchDB and passed to the native VM for execution. Like the OpenWhisk Invoker, the shim process caches function code in-memory to avoid unnecessary calls to the remote database. 

While the shim-layer gives us a number of engineering advantages it also introduces a performance penalty due to the additional network hop between the Linux process and the SEUSS VM. The internal messaging system, part of the EbbRT framework, sends each invocation request as a separate message over an established TCP connection. No coalescing is done to optimize message throughput. We explore the penalties of this design in more detail in our evaluations (\S~\ref{ch:evaluation:macrobenchmarks:performance}).

\section{Evaluation Micro Benchmarks: Function Deployment Techniques}
\label{ch:evaluation:microbenchmarks}
To reiterate, the purpose of SEUSS is to generalize the set of workloads FaaS systems can support by introducing elasticity at the OS level. In this section, we consider operational overheads of these various techniques. We demonstrate that SEUSS supports radically faster function start times, especially in the uncached function case. This is how SEUSS enables elastic workloads. Further, we argue SEUSS radically improves cache density, making SEUSS favorable in the common case of repeat function execution.

We perform a set of micro benchmarks that evaluate the SEUSS primitive---\ecs\ from \snaps\ of Rumprun unikernels---and compare it with the deployment techniques of  Linux processes, Docker containers, and light-weight virtual machines. We conclude this evaluation with a discussion of the upper and lower bounds of the invocation paths for deploying a serverless function SEUSS versus Linux containers.

\subsection{Methodology}
We target a full-featured Node.js runtime environment that is deployed with the invocation driver used in OpenWhisk. We first record the maximum number of Node.js environment instances our Linux VM node can support by sequentially deploying instances until the memory of the VM is saturated.   Table ~\ref{table:memory_footprints} presents the per-node density of Node.js environments deployed using various deployment types on Linux, along with the time to saturation the VM when deploying instances in parallel across all 16 cores.  For system software, we use Ubuntu \textit{18.04} LTS (Bionic Beaver), Linux kernel \textit{v4.15.0}, Docker \textit{v18.09}. 

\begin{table}[t]
  \centering
  \resizebox{\columnwidth}{!}{%
\begin{tabular}{|l|l|l|}
\hline
  \textbf{Node.js Deployment}               & \textbf{Saturation Time} &
  \textbf{Instance Density} \\ \hline
Process (Linux)                   & 93 s                            & 4200                         \\ \hline
Container (Docker/overlay2)          & 569 s                               & 3000                         \\ \hline
microVM (Firecracker)                & 340 s & 450                          \\ \hline
SEUSS unikernel (Rumprun)           & - & 52000                         \\ \hline
\end{tabular}
  }
 \caption{ Density limit and parallel saturation time for the Nodejs runtime
  environments deployed using various isolation techniques on a 88GB
  16 CPUS Linux VM. The SEUSS unikernel result is measured using SEUSS OS.}
  \label{table:memory_footprints}
\vspace{-0.10in}
\end{table}

\subsection{Caching using standard techniques on Linux}
\label{ch:evaluation:microbenchmarks:containers}
We start by evaluating standard isolation technologies that can be used on Linux to deploy and cache function execution state.

\subsubsection{Processes (insufficient encapsulation)}
Processes are not a sufficient isolation mechanism for running untrusted execution in a multi-tenant cloud environment. The value of this result is that it helps to orient the conversation with an optimistic “best case” on Linux with respect to memory sharing and startup latency.

We are able to deploy $4200$ processes of the Node.js runtime on our VM. In parallel, it took about \SI{93}{\s} to saturate the machine using processes.

\subsubsection{Docker Containers}
Using Docker containers and the overlay2 storage driver, we are able to deploy around $3000$ Node.js container instances before saturating the VM.  The additional memory consumption of container over processes is associated with the per-container filesystem state and the loss of DSO sharing across containerized processes. 

We observe that the creation overheads for an individual container are proportional to the number of total container instances active in the system (an observation that has also been corroborated by others~\cite{myvm}). In particular, the creation time for a single Node.js container increased linearly from \SI{541}{\ms} (with no other containers) to over \SI{1.9}{\s} (with around 3000 containers).  Creation times for containers also suffered relative to the number of parallel creations taking place. In particular, we saw the \textit{minimum} creation time to increase by \SI{60}{\ms} \textit{for every additional concurrent creation taking place}.  In our 16-way parallel creation, the 99\% tail latency for deploying a Node.js container is \SI{4.6}{\s}.

Containers have become ubiquitous because of their value in cloud orchestration platforms, but that does not imply they are the right fit for serverless. These two significant non-scalabilities above, lead us to conclude that containers are not well suited for the highly-parallel, ephemeral, and latency-critical requirements of FaaS systems\footnote{The use of Node.js containers in our Apache OpenWhisk experiments (\S\ref{ch:evaluation:macrobenchmarks})) are subject to the same performance penalties as shown here with additional limitations introduced by the in-kernel virtual LAN emulation (as described in \S\ref{ch:evaluation:macrobenchmarks:cachesize})}.

\subsubsection{Lightweight Virtual Machines}
Recently, there has been a movement away from containers as the means to isolate untrusted cloud workloads towards more secure, hardware-enforced virtualization techniques. Lightweight VMs (aka. \textit{microVMs}), together with a "single-user" build of the Linux kernel are combined to create a hardware-enforced isolation mechanism with performance characteristics similar to those of containers.  

Using the Kata Container runtime together with the \texttt{Firecracker} hypervisor, we deploy the Node.js container isolated within a dedicated "microVM".  Unsurprisingly, the virtual machine (along with its dedicated Linux kernel) resulted in an increase of over $100$ MB to the per-instance memory footprint, thereby limiting the number of cached instances that can be deployed on our VM. In addition, the latency to deploy a single Node.js instance via this method is over 3 seconds, primarily due to the setup procedure which now includes booting the Linux kernel prior to deploying the container and runtime.

\subsubsection{Containers vs Lightweight VMs}

From the perspective of the FaaS platform, the technique used to encapsulate execution state should be fast to deploy should have a light-weight memory footprint (for caching) and must be secure.  From a security perspective, hardware virtualization is superior to Linux containers, as the wide system call interface exposed by containers has a large surface for kernel vulnerabilities~\cite{KollerDan}. However, from our results, the use of light-weight VMs over containers introduces overheads that are detrimental for our target of low-latency, large-scale serverless invocations.

The comparable scalability of containers is more opaque.  The time to construct container instances increased considerably (up to $350\%$) as more containers were added to the system, which suggests that the system would perform worse for new invocations as additional function state becomes cached.  We did not see the same relative increase in creation latency for the VM instances.  However, significantly fewer VM instances are simultaneously deployable in our tests. We believe this difference is caused because the container abstraction is implemented across various subsystems of the (host) Linux kernel.  This makes the operations to create and delete individual container instances both non-parallel and non-scalable.  While the scalability of container creation will most likely improve in time. Note that the comparable operations to add or remove virtual domains (via \texttt{kvm}, in our case) are fundamentally less intertwined with the various internal structures of host kernel, and thus are likely to remain both faster and more scalable.  In conclusion, it is not the isolation technology of the light-weight VM that is the bottleneck, but the general-purpose kernel that is deployed within it~\cite{myvm}.

\subsection{Caching using SEUSS}
\label{ch:evaluation:microbenchmarks:seuss}

We measure the time to deploy execution from SEUSS \snaps\ and \ecs\ of a Node.js Rumprun unikernel (Table ~\ref{table:unikernel_size}).  In a saturation test, we are able to deploy over $52,000$ individual Node.js \ec\ instances, each internally blocked on a port with a unique mapping within our NAT layer that allows it to be scheduled and referenced.  This extreme density is enabled by the low memory footprints of the deployed \ec\ instances, which are highly redundant and therefore can get major advantage from COW page sharing (\S\ref{ch:implementation:memory}).

Next, we evaluate the time to deploy, interpret and execute a minimal JavaScript NOP function (i.e., code which simply returns `\texttt{true}`) on SEUSS OS.  For this NOP function, we measure the end-to-end invocation latency from the moment the invocation request is received by a SEUSS OS worker core to the moment the function has finished executing and the result is returned to the worker core. While clearly a microbenchmark, measuring the NOP provides the clearest picture of system-induced overheads. 

The bottom of Table ~\ref{table:unikernel_size} contains the measurements for deploying the NOP function via the hot, warm and cold invocation paths.  The cold and warm paths include deploying the \ec\ from a \snap, while the hot path uses an existing \ec\ which sits idle on that core.  The latency includes the time to set up a TCP connection between the worker core and the invocation driver as well as the time to pass in the invocation arguments.  In addition, the cold start path includes the time to pass in and compile the function code as well as the time to capture the warm start snapshot.

The cold start invocation begins from a Node.js runtime snapshot (captured once during system boot\footnote{This amortizes the \SI{270}{\ms} required to bring up the unikernel Nodejs and the invocation driver, permanently removing it from all invocation paths}), resulting in an end-to-end invocation latency of \SI{7.67}{\ms}\footnote{Interpretation time is minimal for the NOP function, but is likely to be much larger in practice}.  Similarly, warm start invocations begin from an existing function-specific snapshot. Since warm starts do not require function code to be passed in and interpreted, invocation overheads are only \SI{2.95}{\ms}.  
The hot start invocation, which calls into a blocked \ec\ takes only \SI{0.82}{\ms} to execute and return. This optimal path removes \ec\ creation from the execution path, but this is not the main savings. The real benefit here is that the user code has already run once. Initialization has been done and the state has accumulated from user code, kernel paths, and perhaps most importantly, JIT optimizations. Note it is considered safe to re-execute user functions in this way because of the “stateless” assumption of the serverless model. A similar optimization is used with containers to avoid expensive pause/unpause operations.

\begin{table}[]
  \resizebox{\columnwidth}{!}{%
\begin{tabular}{lrll}
  \multicolumn{2}{l}{\textbf{Rumprun Unikernel}} & \textbf{Boot time (ms)} &
  \textbf{Snapshot size } \\ \hline
  \multicolumn{2}{l}{Hello World}                  & 9.2           & 2.1 MB               \\ \hline
  \multicolumn{2}{l}{Python Hello World}            & 393.3         & 10.8 MB                  \\ \hline
  \multicolumn{2}{l}{Python Invocation Driver}  & 1072.3         & 31.8 MB             \\ \hline 
  \multicolumn{2}{l}{Nodejs Hello World}               & 123.9          & 103.9 MB             \\ \hline
  \multicolumn{2}{l}{Nodejs Invocation Driver} & 279.6          & 114.5 MB \\ 
  \multicolumn{2}{l}{} & \textbf{Runtime (ms)}          & \textbf{Allocated Pages} \\  \cline{3-4}  
  & \textit{hot start}:  & 0.82                    & 13 \\ \cline{3-4}    
  & \textit{warm start}: & 2.95                  & 391\\ \cline{3-4}     
  & \textit{cold start}: & 7.67                  & 527\\ 
  \hline
\end{tabular}
  }
  \caption{Boot times and snapshot size taken after boot of various Rumprun unikernels
  deployed on SEUSS (top). Runtime and page
  allocation averages across 475 invocations of a NOP JavaScript function (bottom).}
\label{table:unikernel_size}
\vspace{-0.10in}
\end{table}


\subsubsection{Anticipatory Optimization}
Through a series of \textit{anticipatory optimizations}, we were able to
significantly decrease invocation latencies and memory footprints to what is shown at the bottom of Table ~\ref{table:unikernel_size}.  Prior to making these optimizations, our results were $3$-$4$ times slower. For example, warm start invocation took \SI{13}{\ms} opposed to \SI{3}{\ms}, and a cold start took \SI{40}{\ms} opposed to \SI{8}{\ms}.

The observation was that by sending, compiling and executing an empty JavaScript file prior to capturing the first (runtime) snapshot, we are able to effectively capture into the snapshot a "warmed up" version of, for example, the unikernel network stack, as well as the interpreter’s compilation and execution paths. As a result, the initial runtime snapshot is considerably larger, but each of the function-specific snapshots has decreased in size. Note that in our prototype,   machines holds a single copy of each runtime snapshots, while every one of the possibly tens of thousands of function-specific snapshots can benefit from memory and latency savings.
 In our evaluation of OpenWhisk with a SEUSS node, the smaller snapshot sizes results in a doubling of our cache size---from $16,000$ to $32,000$ NOP functions (\S\ref{ch:evaluation:macrobenchmarks:utilization})---during a high density throughput test.

\subsubsection{Rapid Deployment in SEUSS vs Linux}

\begin{figure}[h]
\centering
\includegraphics[width=1.1\linewidth]{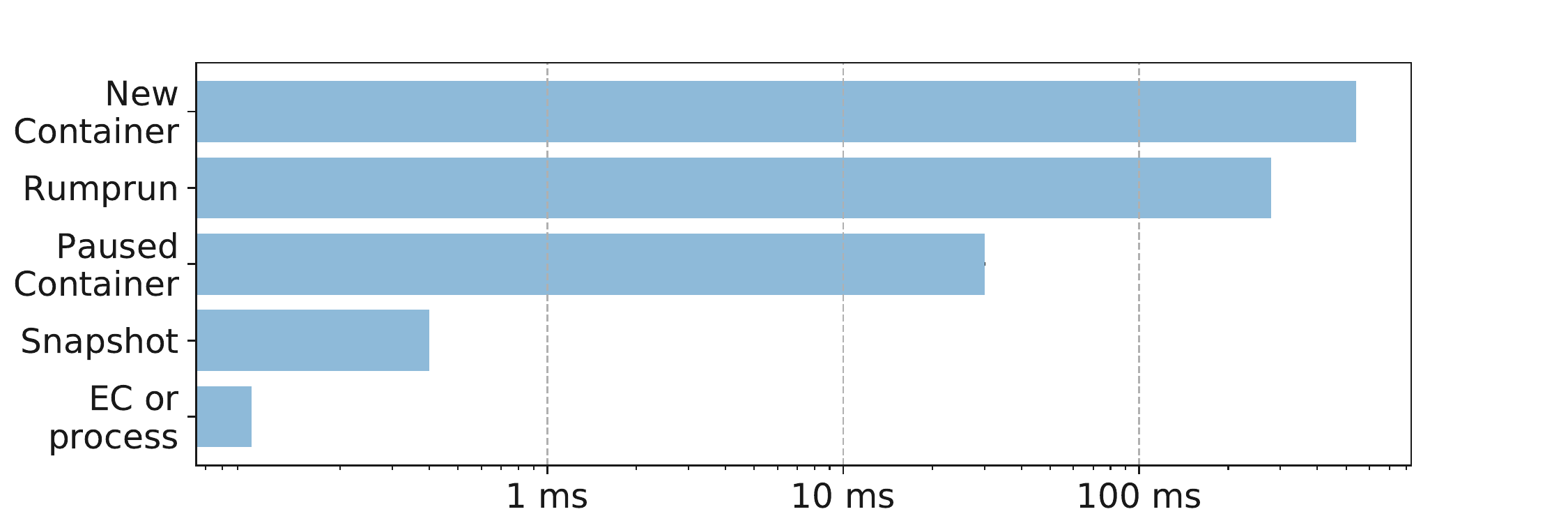}
\caption{Start times for different execution environments. X-axis is log scale
  (ms)}
\label{fig:start_times}
\vspace{-0.10in}
\end{figure}

To conclude these micro benchmark evaluations, we discuss the expected best and worst overheads across the two deployment strategies we employ in our evaluation of OpenWhisk, SEUSS \snaps\ vs Docker containers.  Figure~\ref{fig:start_times} presents a summary of the various deployment overheads (graphed on a log scale) for the Node.js runtime evaluated throughout this evaluation.  

\begin{description}[leftmargin=1em]
\item [In the most optimal case,] 
function execution can begin in just a few hundred microseconds. This can be achieved on both SEUSS OS and Linux by a \ec\ or \textit{unpaused} container that has been fully initialized to the particular function.  As shown in \S\ref{ch:evaluation:macrobenchmarks:utilization}, this optimal case is more easily achievable in SEUSS OS across different function invocations due to the increased cache density.

\item [The next best option] 
  for Linux, is the use of a pre-initialized environment cached inside of a \textit{paused} container. In our evaluations, the time to unpause a Node.js Docker container took between \SI{30}{\ms} and \SI{50}{\ms} depending on the level of concurrency.  In comparison, SEUSS can deploy a similarly-initialized \ecs\ from a snapshot in about \SI{400}{\us}. In addition, our technique of deploying from a snapshot is far more flexible as multiple \ecs\ can be deployed in parallel from a single shared snapshot (while a paused container is consumed as part of the invocation).

\item [The worst-case] 
  on Linux requires a new container instance to be constructed and
    initialized, which we observe takes between \SI{540}{\ms} to \SI{4.6}{\s}
    (not including the time to import and interpret the function code and
    arguments). Comparably, the worst-case on SEUSS OS is to deploy an initialized runtime from a snapshot prior to importing the function code, which takes around \SI{400}{\us}. Note that without snapshotting, the SEUSS cold path is comparable to container creation Figure~\ref{fig:start_times} “Rumprun” bar. The runtime snapshot permanently removes this cost from all paths.
  
\end{description}

\section{Evaluation Macro Benchmarks: FaaS Platform Performance}
\label{ch:evaluation_openwhisk}
\label{ch:evaluation:macrobenchmarks}
The previous micro benchmark evaluation (\S\ref{7}) provides intuition for where we can expect to see benefit from the SEUSS approach when applied to optimize invocations within a FaaS platform. Namely in the radically faster function deployment path (especially for cold starts), and the dramatically denser function caches.

In this second evaluation, we analyze the performance of requests to a multi-node FaaS cluster (Apache OpenWhisk) with our SEUSS kernel prototype used as a drop-in replacement for Linux on a virtual compute node.  Using a custom benchmark tool that we developed, we evaluate the platform performance across a variety of invocation patterns and function types. The following highlights the primary results:
\begin{enumerate} 
\item In the face of increasing diverse invocation patterns, the SEUSS backend is able to sustain high-throughput long after Linux node a performance wall due to cache thrashing. (\S\ref{ch:evaluation:macrobenchmarks:utilization})
\item SEUSS radically outperforms Linux when it comes to deploying new environments for uncached functions, but also cached functions are processed faster as Linux quickly get bogged down with cache invalidations and container construction overheads(i.e., when many different functions are being invoked).
\item SEUSS can support large-scale \textit{request burst} that fully overwhelm the Linux backend. (\S\ref{ch:evaluation:macrobenchmarks:burst})
\item SEUSS prototype provides a scalable solution to multiplexing functioning access to the external network. However, in our existing implementation, stressing the external networking ingress and egress paths show non-scalabilities that limit the network throughput of the function when compared to Linux containers. (We believe this can be addressed through implementation and its not inherent to the SEUSS model).
\end{enumerate}

\subsection{Methodology}
We run our evaluations on a four-node cluster of physical machines connected via a 10GbE network on a private VLAN through a commodity switch. Each machine contains two 8-core Intel Xeon E5-2660 processors (16 cores in total) running at $2.20$ GHz, $132$ GB of RAM, and a Solarflare Communications SFC9120 Ethernet card. The processors have been configured to disable Turbo Boost, hyper-threads, and dynamic frequency scaling. For software, we use Ubuntu \textit{18.04} LTS (Bionic Beaver), Linux kernel  \textit{v4.15.0}, Docker \textit{v18.09}, and OpenWhisk \textit{v0.9.0}.

We dedicate two of the four physical machines to host OpenWhisk, one machine to host the benchmark, and one machine to hosts an external HTTP server (used in \S\ref{ch:evaluation:macrobenchmarks:burst}).  Of the two OpenWhisk machines, the first hosts the control plane components (i.e., the platform controller, API server, message service, and internal databases) deployed within containers on the Linux host.  The second OpenWhisk machine hosts a single \texttt{qemu-kvm} VM instance which acts as the OpenWhisk compute node across all experiments. We use a VM to maintain identical hardware specifications for both the SEUSS OS and Linux compute node. The VM is configured with $16$ VCPUs, $88$ GB of memory, a virtio/vhost paravirtualized network device, and an in-memory (ramdisk) filesystem. 

\subsubsection{Benchmark}
For these evaluations, we've built a custom OpenWhisk load generation benchmark~\cite{}. The benchmark works by sending a series of parallel, synchronous invocation requests to the OpenWhisk API gateway. For each request, we record the round-trip time for OpenWhisk to process the request, run the function, and return the result to the benchmark.

Each trial of the benchmark has three main parameters: Invocation Count ($N$), Unique Function Set Size ($M$), and Concurrent Requests ($C$).  The benchmark then generates a work queue of $N$ invocation requests across the set of $M$ functions in a random order\footnote{For a fair comparison, the random ordering of invocations is pre-determined and held consistent across runs of the benchmark systems}.  The evaluation trial involves $C$ worker threads pulling invocation requests (one at a time) from the shared queue and starting a new HTTP connection with the API server to process the request\footnote{The maximum number of requests in flight at a given time is $C$}. The benchmark reports the RTT latency for each invocation as well as the aggregate throughput achieved by all $C$ threads. The trial is finished once a response has been received for each of the $N$ requests.

\subsection{Apache OpenWhisk Configuration}
Each trial of the benchmark is performed on a fresh deployment of the OpenWhisk that has been populated with the set of ($M$) user functions run by the benchmark.  We have disabled all platform-enforced quotas and rate limits in OpenWhisk. We also prevent Docker containers from being paused when they are not in use (thereby granting the Linux VM a slight performance boost under high-throughput evaluation).  As a stand-alone evaluation, we measure the control plane overheads of our OpenWhisk cluster by modifying the invocation service to immediately reply with 'success' to each invocation request.   Using a sequential stream of requests, we observed the mean latency for an individual request to be \SI{60}{\ms} with a $99\%$ tail latency of \SI{64}{\ms}. This latency account for the time for the request message to pass from the benchmark, through the OpenWhisk control plane, to the invocation node, and back again. In addition, we observed the control plane throughput to plateau between $32$ and $64$ parallel requests, with a peak throughput of 220 requests per second (without any function executions).

%

\subsubsection{Linux Container Cache}
\label{ch:evaluation:macrobenchmarks:cachesize}
On our Linux compute node, the function state is cached inside of \textit{unpaused} Docker containers, and the max cache size limit is set to $1024$ containers.  For the throughput tests (\S\ref{ch:evaluation:macrobenchmarks:utilization}), we have disabled the OpenWhisks "warm pool" container cache, as we observed that the automatic initialization of containers hurt platform throughput when under heavy load. The warm pool container cache is enabled for the burst experiment (\S\ref{ch:evaluation:macrobenchmarks:burst}). 

The container cache size limit of $1024$ is considerably lower than the max container density of about 3000 instances (Table ~\ref{table:memory_footprints}). We originally set the cache size closer to the observed density limit and found that a majority of invocation requests result in an error. Upon investigation, we observed high rates of dropped packets on the internal network bridge that the platform uses to communicate with the active containers. Due to dropped packets, the connection between the platform and the invocation driver inside the container repeatedly fails and the invocation is eventually aborted.  

Here we witness another limitation in the legacy system being deployed.  The use of virtual Ethernet means that packet processing between the connected endpoints takes place entirely within the Linux kernel. Therefore, a single broadcast (e.g. ARP, DHCP) packet sent over a bridge with $N$ connected endpoints must be processed by the kernel $N$ separate times~\footnote{Parallel packet processing proves problematic given that the Linux kernel has ingress queue length of 1000}. With a max cache size of 1024 containers---the default limit of endpoints on a Linux bridge---we still witness connections failures in our large-scale evaluations (\S~\ref{ch:evaluation:macrobenchmarks:burst}). 

\begin{figure}[h]
\centering
\includegraphics[width=\linewidth]{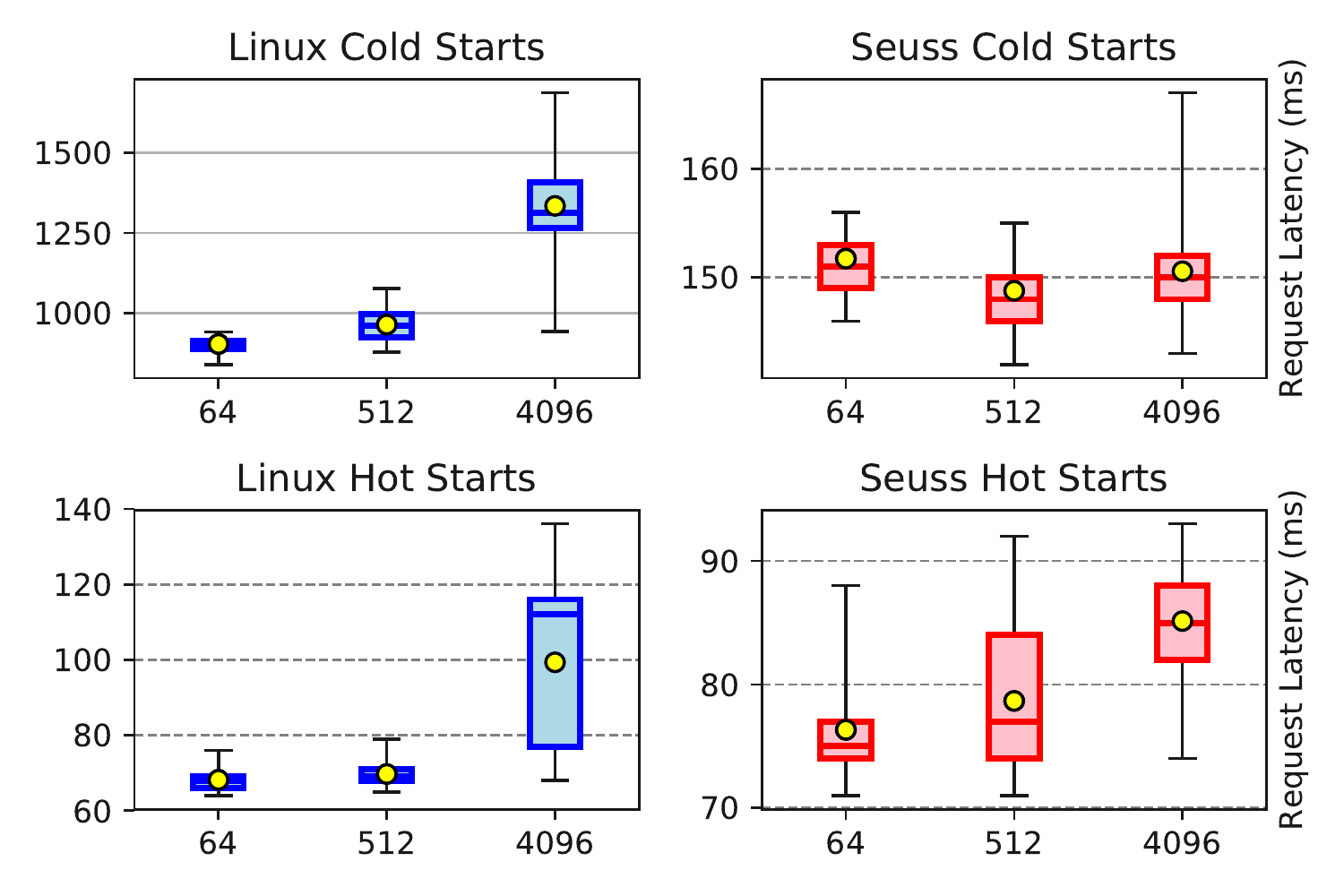}
\caption{
  End-to-end request latencies for a single stream of NOP JavaScript execution across three function set sizes. 1st, 25th, 50th, 75th, 99th percentiles and the mean (dot). 
}
\label{fig:throughput_latencies}
\vspace{-0.10in}
\end{figure}

\begin{figure}[h]
\centering
\includegraphics[width=\linewidth]{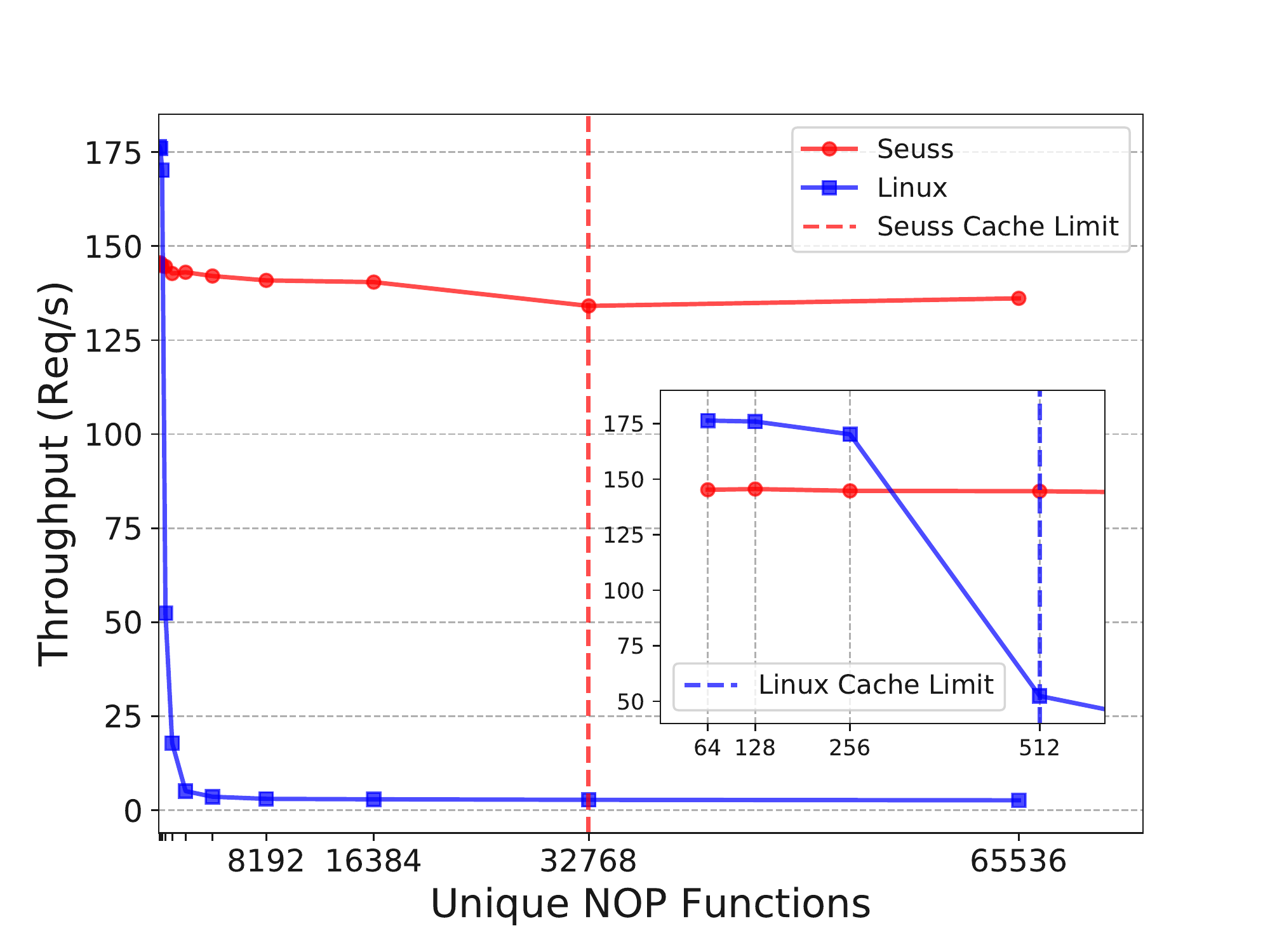}
\caption{OpenWhisk throughput across an increasingly diverse set of invocations}
\label{fig:throughput}
\vspace{-0.10in}
\end{figure}

\subsection{Throughput with Cached / Uncached Invocations}
\label{ch:evaluation:macrobenchmarks:utilization}

This experiment is designed to stress the FaaS platform’s ability to effectively cache and deploy function execution when exposed to increasingly diverse function invocations. To provide "diversity", we increase the set of unique functions that are invoked across each trial, while keeping the total number of invocations consistent.  While each function is logically different, the code being run is an identical JavaScript NOP. We use a NOP function to minimize the time spent in function-specific work, thereby stressing the invocation path overheads of the platform.  


Figure~\ref{fig:throughput} presents the achieved throughput OpenWhisk using the SEUSS OS or Linux compute node.  Trials are represented as points along the x-axis. For each trial, we doubled the number of unique functions that the benchmark invokes (ranging from $64$ to $65536$).  For each trial, the benchmark sends a continuous stream of invocation requests from $32$ threads until the measured throughput reaches a point of stability.

The throughput of the trials with low diversity (the left-most points on the lines) represents a “best-case scenario” for the platform's performance, as the vast majority of requests result in "hot" repeat invocations and the caches remained underutilized. In contrast, as diversity increases past the relative cache limits (the right-most points on the lines) it presents a “worst-case scenario” for the system, as an increasing majority of requests require cold path invocations.  
As shown in Figure~\ref{fig:throughput}, the overall throughput of the FaaS platform depends heavily on the total amount of unique function state that can be efficiently cached at the site of the invocation. For example, throughput on the Linux node drops severely once the container cache is saturated around 512 functions (with 32 request concurrency. SEUSS, being designed for efficient caching, can hold over 40,000 \snaps\ and \ecs of the minimal NOP function before the memory of the VM is exhausted. 

On an all-unique workload (zero repeat invocation), the SEUSS OS has a $52\times$ speedup over Linux.  This primarily dues to the cold invocation paths on SEUSS OS begin from an initialized Node.js runtime snapshot. Conversely, when the cache is saturated on Linux, there is no longer an opportunity to pre-initialize containers for not-yet-seen function invocation. Instead, every cold start requires both cache eviction (container deletion) and new container creation. As demonstrated in \S~ref{ch:evaluation:microbenchmarks:containers}, the performance of container operations suffers proportionally to the container occupancy. The comparable difference in cold path invocation latencies, when the cache is under-utilized (64 functions) and oversaturated (2048 functions), can be seen in Figure~\ref{fig:throughput_latencies}.

In the low-diversity trials, the throughput of Linux is $21\%$ higher than that of SEUSS OS. This is because the Linux node provides better hot start latencies while at low cache utilization (Figure~\ref{fig:throughput_latencies}). In the SEUSS prototype, the hot start pathways are burdened with an additional packet processing and network hop between the SEUSS OS \textit{shim process} and the OpenWhisk control plane(\S~\ref{ch:seuss:implementation:openwhisk}), which adds about \SI{8}{\ms} to a round-trip request latency. We believe this gap can be reduced through simple implementation changes to the SEUSS OS shim process. However, low-diversity performance is not the focus of the SEUSS OS design.

\subsection{Platform Resiliency to Bursty Workloads}
\label{ch:evaluation:macrobenchmarks:burst}
In this experiment, we aim to stress the platform capacity for caching and invocation when exposed to the sudden arrival of parallel \textit{bursts} of invocation requests.  The experiment involves giving the system a moderate level of "background" request load and then hitting it with repeat large-scale batches of parallel requests. Instead of NOP functions, this experiment runs a combination of CPU-heavy functions (to model a computational workload) and a mixture of functions blocked on external IO (representing a more “normal” background behavior). 

\subsubsection{Experiment Setup}
The background utilization consists of a continuous stream of (hot start) invocations made to a set of IO-bound functions. Each function makes an external network call to a remote \texttt{HTTP} server, which waits for \SI{250}{\ms} before sending an \texttt{OK} reply (the function finishes once the reply is received).  To generate the background utilization, we configure the benchmark to use $128$ threads and make requests to a total of 16 unique IO functions. The benchmark is rate-throttled to a limit of 72 requests per second (about $40\%$ the observed throughput of the platform. On Linux, around $40\%$ of the container cache is consumed by the background stream of requests.

On top of the background stream, we issue a series of invocation bursts, each consists of $128$ invocation requests sent in parallel to a unique never-been-seen function.  Bursts are sent at a fixed frequency of 32, 16, or 8 seconds between bursts, and we observe if the background load is affected and if the system is successful in handling all the parallel requests.  All functions in the bursts do CPU-bound computation that takes around \SI{150}{\ms} to complete. Invocations within a single burst are to the same function, while functions across bursts are unique.  presents the results.

\begin{figure}[h]
  \includegraphics[width=\linewidth]{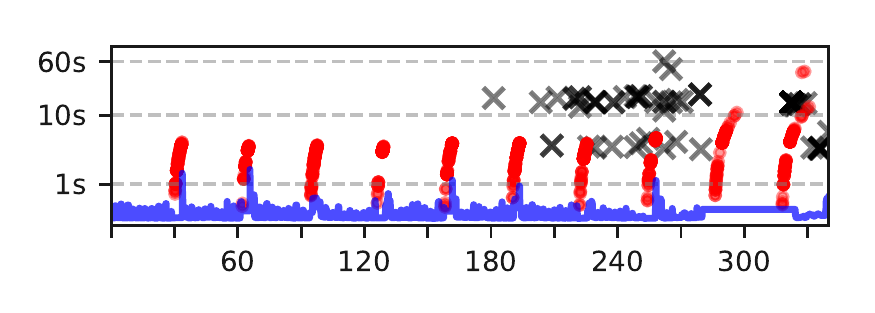} \vspace{-0.45in} \\
  \includegraphics[width=\linewidth]{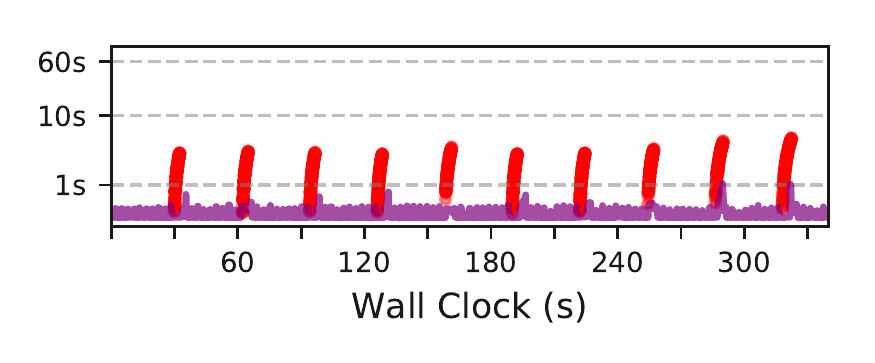}  
  \vspace{-0.10in}
  \caption{32s bursts. Linux (top) and SEUSS (bottom). Request latency (log scale) is
  measured on the y axis, trial wall clock time on the x axis
    (\SI{340}{\s} total). Plotted points represents
    a successful completion with the x being the time the request was sent, and y
    being the RRT latency. Failed requests are marked with an '\texttt{x}'}
  \label{fig:burst_1}
  \vspace{-0.10in}
\end{figure}

\subsubsection{Burst Resiliency on Linux }
On Linux, we configure OpenWhisk to maintain a cache of 256 “warm start” containers of the Node.js runtime. This container cache provides warm path invocations for the sudden bursts of never-been-seen invocation. However, these pre-initialized containers directly compete with the other containers on the node, affecting both container creation times and the number of "cached" containers that can be used to enable hot starts.  With a frequency of \SI{32}{\s} between bursts (top left graph, Figure~\ref{fig:burst}), the Linux container cache can be repopulated between bursts, and so the sudden arrival of requests is met by warm start invocations.  However, beginning around the 6th burst, the container cache limit is hit and some of the requests begin to return an error (marked in the figures as a black 'x').  When the burst frequency is increased to $16$s and then $8$s (middle right, bottom right graphs, Figure~\ref{fig:burst}), the container cache is not able to repopulate between bursts. In these cases, only the first burst sees the warm start invocations, while all additional burst see cold start overheads (of up to \SI{10}{\s}) and failed requests. In all cases, the reliability of the Linux node suffers as the container cache becomes saturated, resulting in a majority of request failures after only a few bursts.  When the bursts continue after the container cache is saturation OpenWhisk stop processing request all together for a number of minutes, presumably in a recovery mode.  

\begin{figure}[h]
  \includegraphics[width=\linewidth]{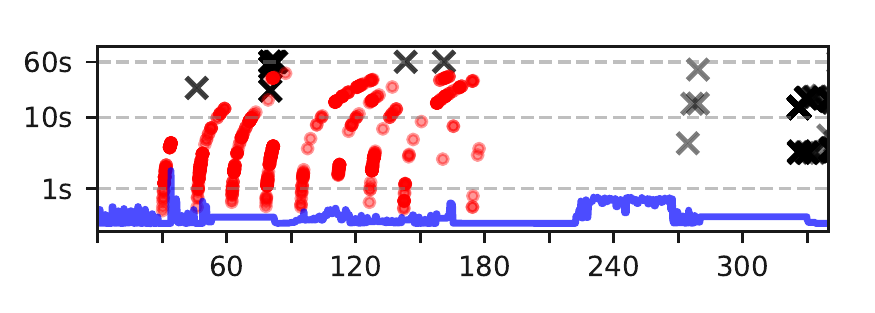} \vspace{-0.45in} \\
  \includegraphics[width=\linewidth]{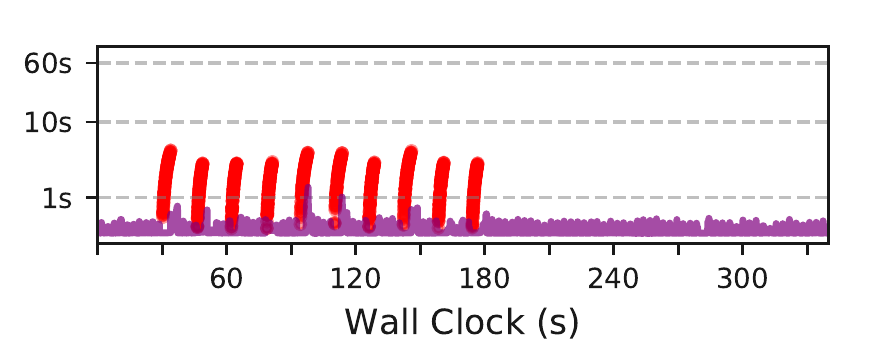}  
  \caption{16s bursts. Linux (top) and SEUSS (bottom)}
  \vspace{-0.10in}
  \label{fig:burst_2}
  \vspace{-0.10in}
\end{figure}

\subsubsection{Burst Resiliency on SEUSS OS}
\begin{figure}[h]
  \includegraphics[width=\linewidth]{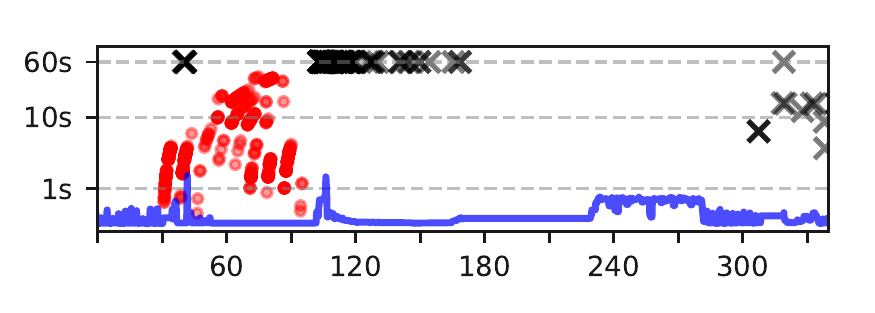} \vspace{-0.45in} \\
  \includegraphics[width=\linewidth]{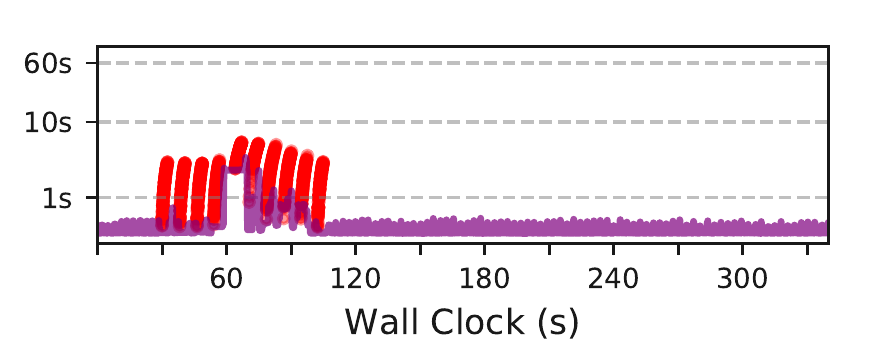}  
  \caption{8s bursts. Linux (top) and SEUSS (bottom)}
  \vspace{-0.10in}
  \label{fig:burst_3}
  \vspace{-0.10in}
\end{figure}

The same experiment of OpenWhisk with a SEUSS OS node presents a stark difference in results. With SEUSS OS, OpenWhisk is able to handle every request we send (no request return an error to our benchmark). Furthermore, as every new invocation is handled from a snapshot, the request latencies remain consistently low across all three burst frequencies. Only at the highest frequency (8 seconds between bursts) does the background stream become disturbed (Figure~\ref{fig:burst_3}). This is because IO tasks get queued behind  CPU tasks which are run to completion. This is not inherent to the design, there is no known significant benefit SEUSS derives from this policy; addressing it with scheduling would be a natural step for future work. Finally, as each burst adds only one additional snapshot to the cache, we would have to process many thousands of additional bursts on SEUSS OS before the cache would be full. Note the difference in the Linux approach where each truly parallel execution of each function requires its own container.

\subsection{External TCP/IP Performance}
\label{ch:evaluation:macrobenchmarks:performance}
One shortcoming of our system design is the processing overheads required to masquerade external TCP/IP traffic in and out of the running \ecs (\S\ref{ch:implementation:networking}). To evaluate the impact of these overheads, we deploy an I/O-heavy JavaScript function that reads a 1KB, 10KB or 100KB payload from an external HTTP server. We deploy the function repeatedly on an otherwise quiet deployment of OpenWhisk and measure the average RTT latencies. For the 1KB and 10KB payloads, the function took about 20\% longer using the SEUSS OS compute node. When increasing the payload size to 100KB, the network throughput was an order of magnitude better on Linux.  

Unsurprising to us, these results are symptomatic of one of our major design decisions, to deploy general-use unikernels in user-space, and to multiplex network traffic at layer-3 (effectively doubling the processing overhead per-packet). In a sense, this is an old problem in the world of virtualization, to which multiple hardware and software solutions have been presented. However, in our case, a simple and scalable solution to this design puzzle remains an ongoing topic of debate among our group. 

\section{Conclusion}
\label{ch:seuss:conclusion}

As evidenced by our evaluations, modern FaaS platforms achieve fast deployments when they reuse cached execution state of previously run functions but suffer disproportionately when no cached state is available. In SEUSS, we improve on the approach of caching by deploying execution from \snaps, which enables function environments to be created faster and more environments can be cached on a node. 

Deploying execution from \snaps\ in SEUSS benefits both hot and cold invocation paths. Cold starts have a unique advantage on SEUSS as they can deploy from a reusable runtime snapshot. In our results, cold start overheads for deploying a NOP function from a snapshot is less than 8ms. On Linux, the cold start overheads begin with a 500ms to 4s overhead to deploy a new containerized environment (\S\ref{ch:evaluation:microbenchmarks:containers}).  

For hot starts, SEUSS’ higher-density cache allows a FaaS compute node to improve the probability of a hit. In our results, we demonstrate that a SEUSS node can cache at least an order of magnitude more isolated executing environments than Docker containers and more than two orders magnitude than with light-weight VMs (\S\ref{ch:evaluation:microbenchmarks:snapshots}). 

By combining the benefits of low latency cold starts and simplified immutable memory caching, SEUSS runs high-demand \textit{bursty} workloads that are currently unsupported by the traditional approach (\S\ref{ch:evaluation:macrobenchmarks:burst}).  

Our results show that the serverless function need no longer be considered a heavyweight computational primitive. When parallel executions can be deployed in milliseconds (from snapshots which can be created even faster), the serverless function becomes an intuitive way to enact computational elasticity at arbitrary scales. With the function as a performant base primitive, it will be possible for FaaS platforms to unlock a long-promised mutualism: users get instantaneous access to on-demand parallelism, while providers manage a much finer-grained resource bin packing problem, trading hundreds of VMs for tens of thousands of functions. SEUSS demonstrates that it is possible to bump FaaS out of its current computational niche by introducing elasticity at the OS level.

\pagestyle{empty}
\clearpage
\bibliographystyle{plain}
\bibliography{biblio}
\pagestyle{empty}

\end{document}